\documentclass[prl,preprintnumbers,amsmath,amssymb]{revtex4}

\usepackage{mathrsfs}
\usepackage{graphicx}
\usepackage{dcolumn}
\usepackage{bm}

\begin{document}

\title{Backward transfer entropy: Informational measure for detecting hidden Markov models and its interpretations in thermodynamics, gambling and causality}

\author{Sosuke Ito$^{1,2}$}
 
\affiliation{$^1$Department of Physics, Tokyo Institute of Technology, Oh-okayama 2-12-1, Meguro-ku, Tokyo 152-8551, Japan\\
$^2$FOM Institute AMOLF, Science Park 104, 1098 XG Amsterdam, The Netherlands\\
sosuke@stat.phys.titech.ac.jp}
\maketitle

{\bf ABSTRACT}

The transfer entropy is a well-established measure of information flow, which quantifies directed influence between two stochastic time series and has been shown to be useful in a variety fields of science. Here we introduce the transfer entropy of the backward time series called the backward transfer entropy, and show that the backward transfer entropy quantifies how far it is from dynamics to a hidden Markov model. Furthermore, we discuss physical interpretations of the backward transfer entropy in completely different settings of thermodynamics for information processing and the gambling with side information. In both settings of thermodynamics and the gambling, the backward transfer entropy characterizes a possible loss of some benefit, where the conventional transfer entropy characterizes a possible benefit. Our result implies the deep connection between thermodynamics and the gambling in the presence of information flow, and that the backward transfer entropy would be useful as a novel measure of information flow in nonequilibrium thermodynamics, biochemical sciences, economics and statistics.

{\bf Introduction}

	In many scientific problems, we consider directed influence between two component parts of complex system. To extract meaningful influence between components parts, the methods of time series analysis have been widely used~\cite{Hamilton, Ahmed, Gao}. Especially, time series analysis based on information theory~\cite{Cover-Thomas} provides useful methods for detecting the directed influence between component parts. For example, the transfer entropy (TE)~\cite{Schreiber, Schreiber2, HlavackovaVejmelka} is one of the most influential informational methods to detect directed influence between two stochastic time series. The main idea behind TE is that, by conditioning on the history of one time series, informational measure of correlation between two time series represents the  information flow that is actually transferred at the present time.	
	 Transfer entropy has been well adopted in a variety of research areas such as economics~\cite{Marschinski}, neural networks~\cite{Lungerella, Vicente,  Wibral2},  biochemical physics~\cite{Bauer, ItoNatcom, HartichSeifertSensor} and statistical physics~\cite{ItoPRL, HartichSeifert, Prokopenko1, Prokopenko2, Barnett}. Several efforts to improve the measure of TE have also been done~\cite{Staniek, Wibral, Williams}.
	 
	In a variety of fields, a similar concept of TE has been discussed for a long time.  In economics, the statistical hypothesis test called as the Granger causality (GC) has been used to detect the causal relationship between two time series~\cite{Grangerbooks, Granger}. Indeed, for Gaussian variables, the statement of GC is equivalent to TE~\cite{BarnettSeth}. In information theory, nearly the same informational measure of information flow called the directed information (DI)~\cite{Massey, Marko} has been discussed as a fundamental bound of the noisy channel coding under causal feedback loop. As in the case of GC, DI can be applied to an economic situation~\cite{Permuter,Hidaka}, that is the gambling with side information~\cite{Cover-Thomas, Kelly}.

In recent studies of a thermodynamic model implementing the Maxwell's demon~\cite{review,SagawaUeda}, which reduces the entropy change in a small subsystem by using information, TE has attracted much attention~\cite{ItoNatcom, AllahverdyanMahler, ItoPRL, HartichSeifert, HorowitzEsposito, HorowitzSandberg, Horowitz, ItoBook, HartichSeifertSensor, Springer}. In this context, the transfer entropy from a small subsystem to other systems generally gives a lower bound of the entropy change in a subsystem~\cite{AllahverdyanMahler, ItoPRL, HartichSeifert}. As a tighter bound of the entropy change for Markov jump process, another directed informational measure called the dynamic information flow (DIF)~\cite{HorowitzEsposito} has also been discussed~\cite{AllahverdyanMahler, HorowitzEsposito, HorowitzSandberg, Horowitz, ItoBook, Cafaro,ShiraishiSagawa, ShiraishiItoKawaguchiSagawa, Rosinberg, YamamotoIto, Springer}.

In this article, we provide the unified perspective on different measures of information flow, i.e., TE, DI, and DIF. To introduce TE for backward time series~\cite{ItoNatcom, Springer}, called  {\it backward transfer entropy} (BTE), we clarify the relationship between these informational measures. By considering BTE, we also obtain a tighter bound of the entropy change in a small subsystem even for non Markov process. In the context of time series analysis, this BTE has a proper meaning: an informational measure for detecting a hidden Markov model. From the view point of the statistical hypothesis test, BTE quantifies an anti-causal prediction. These fact implies that BTE would be a useful directed measure of information flow as well as TE.

Furthermore, we also discuss the analogy between thermodynamics for a small system~\cite{review,Sekimoto,Seifert} and the gambling with side information~\cite{Cover-Thomas, Kelly}. To considering its analogy, we found that TE and BTE play similar roles in both settings of thermodynamics and gambling: BTE quantifies a loss of some benefit while TE quantifies some benefit. Our result reveals the deep connection between two different fields of science, thermodynamics and gambling. 

{\bf Results}

{\bf Setting.} 
We consider stochastic dynamics of interacting systems ${\mathcal X}$ and ${\mathcal Y}$, which are not necessarily Markov processes.  We consider a discrete time $k$ $(= 1, \dots, N)$, and write the state of ${\mathcal X}$ (${\mathcal Y}$) at time $k$ as $x_k$ ($y_k$). Let $x_k^{(l)} := \{ x_k, \dots, x_{k-l+1} \}$ ($y_k^{(l)} := \{ y_k, \dots, y_{k-l+1} \}$) be the path of system ${\mathcal X}$ (${\mathcal Y}$) from time $k-l+1$ to $k$ where $l \geq 1$ is the length of the path. The probability distribution of the composite system at time $k$ is represented by $p(X_k= x_k, Y_k=y_k)$, and that of paths is represented by $p(X_{k}^{(k)} = x_k^{(k)}, Y_{k}^{(k)} =y_{k}^{(k)})$, where capital letters (e.g., $X_k$) represent random variables of its states (e.g., $x_k$).

The dynamics of composite system are characterized by the conditional probability $p(X_{k+1}=x_{k+1}, Y_{k+1}=y_{k+1} |X_{k}^{(k)} =x_k^{(k)}, Y_{k}^{(k)} =y_{k}^{(k)} )$ such that 
\begin{eqnarray}
&&p(X_{k+1}^{(k+1)}=x_{k+1}^{(k+1)}, Y_{k+1}^{(k+1)}=y_{k+1}^{(k+1)})  \nonumber \\
&&= p(X_{k+1}=x_{k+1}, Y_{k+1}=y_{k+1} |X_{k}^{(k)} = x_k^{(k)},Y_{k}^{(k)} = y_{k}^{(k)} ) p(X_{k}^{(k)} = x_k^{(k)},Y_{k}^{(k)} = y_{k}^{(k)}),
\end{eqnarray}
where $p(A=a|B=b) := p(A=a, B=b)/ p(B=b)$ is the conditional probability of $a$ under the condition of $b$.

{\bf Transfer entropy.} 
Here, we introduce conventional TE as a measure of directed information flow, which is defined as the conditional mutual information~\cite{Cover-Thomas} between two time series under the condition of the one's past. The mutual information characterizes the static correlation between two systems. The mutual information between $X$ and $Y$ at time $k$ is defined as
\begin{equation}
I (X_k;Y_k) := \sum_{x_k, y_k} p(X_k= x_k, Y_k= y_k) \ln \frac{p(X_k=x_k, Y_k=y_k)}{p(X_k=x_k)p(Y_k=y_k)}.
\end{equation}
This mutual information is nonnegative quantity, and vanishes if and only if $x_k$ and $y_k$ are statistically independent (i.e., $p(X_k=x_k, Y_k=y_k)=p(X_k=x_k)p(Y_k=y_k)$)~\cite{Cover-Thomas}. This mutual information quantifies how much the state of $y_k$ includes the information about $x_k$, or equivalently the state of $x_k$ includes the information about $y_k$.
In a same way, the mutual information between two paths $x_{k}^{(l)}$ and $y_{k'}^{(l')}$ is also defined as
\begin{equation}
I (X_{k}^{(l)}; Y_{k'}^{(l')}) := \sum_{x_{k}^{(l)}, y_{k'}^{(l')}} p(X_{k}^{(l)}= x_{k}^{(l)}, Y_{k'}^{(l')}=y_{k'}^{(l')}) \ln \frac{p(X_{k}^{(l)}= x_{k}^{(l)}, Y_{k'}^{(l')}=y_{k'}^{(l')})}{p(X_{k}^{(l)}=x_{k}^{(l)})p( Y_{k'}^{(l')}=y_{k'}^{(l')})}.
\end{equation}
While the mutual information is very useful in a variety fields of science~\cite{Cover-Thomas}, it only represents statistical correlation between two systems in a symmetric way.  In order to characterize the directed information flow from $X$ to $Y$, Schreiber~\cite{Schreiber} introduced TE defined as
\begin{equation}
T_{X_{k}^{(l)} \to Y_{k'+1}^{(l'+1)}} :=I(X_{k}^{(l)} ; Y_{k'+1}^{(l'+1)}) -I(X_{k}^{(l)}; Y_{k'}^{(l')}),
\label{transfer}
\end{equation}
with $k \leq k'$. Equation~(\ref{transfer}) implies that TE $T_{X_{k}^{(l)} \to Y_{k'+1}^{(l'+1)}}$ is an informational difference about the path of the system ${\mathcal X}$ that is newly obtained by the path of the system ${\mathcal Y}$ from time $k'$ to $k'+1$. Thus, TE $T_{X_{k}^{(l)} \to Y_{k'+1}^{(l'+1)}}$ can be regarded as a directed information flow from ${\mathcal X}$ to ${\mathcal Y}$ at time $k'$. This TE can be rewritten as the conditional mutual information~\cite{Cover-Thomas} between the paths of $\mathcal{X}$ and the state of $\mathcal{Y}$ under the condition of the history of $\mathcal{Y}$:
\begin{align}
T_{X_{k}^{(l)} \to Y_{k'+1}^{(l'+1)}}  &= I(X_{k}^{(l)} ;Y_{k'+1}| Y_{k'}^{(l')} )  \nonumber \\
&:=  \sum_{x_{k}^{(l)} , y_{k'+1}^{(l'+1)}} p(X_{k}^{(l)}=x_{k}^{(l)} , Y_{k'+1}^{(l'+1)}=y_{k'+1}^{(l'+1)}) \ln \frac{p(Y_{k'+1}= y_{k'+1}|X_k^{(l)}=x_{k}^{(l)}, Y_{k'}^{(l')}= y_{k'}^{(l')})}{p(Y_{k'+1}= y_{k'+1}|Y_{k'}^{(l')}=y_{k'}^{(l')})},
\end{align}
which implies that TE is nonnegative quantity, and vanishes if and only if the transition probability in ${\mathcal Y}$ from $y_{k'}^{(l')}$ to $y_{k'+1}$ does not depend on the time series $x_{k}^{(l)}$, i.e., $p(Y_{k'+1}= y_{k'+1}|X_k^{(l)}=x_{k}^{(l)}, Y_{k'}^{(l')}= y_{k'}^{(l')})=p(Y_{k'+1}= y_{k'+1}|Y_{k'}^{(l')}= y_{k'}^{(l')})$). [see also Fig. 1(a)]
\begin {figure}[h]
\centering
\includegraphics[width=7cm]{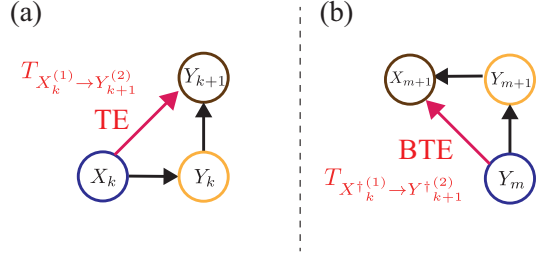}
\caption{Schematics of TE and BTE. Two graphs (a) and (b) are the Bayesian networks corresponding to the joint probabilities $p(X_k=x_k, Y_k=y_k, Y_{k+1}=y_{k+1})= p(X_k=x_k)p(Y_k=y_k|X_k=x_k) p(Y_{k+1}=y_{k+1}|X_k=x_k, Y_k=y_k)$ and $p(X_{m+1}=x_{m+1}, Y_m =y_m, Y_{k+1}=y_{k+1})= p(Y_m=y_m) p(Y_{m+1}=y_{m+1}|Y_m=y_m) p(X_{m+1}=x_{m+1}|Y_{m+1}=y_{m+1}, Y_m=y_m)$, respectively (see also Ref.~\cite{ItoPRL, ItoBook}). (a) Transfer entropy $T_{X_k^{(1)} \to Y_{k+1}^{(2)}}$ corresponds to the edge from $X_k$ to $Y_{k+1}$ on the Bayesian network. If TE $T_{X_k^{(1)} \to Y_{k+1}^{(2)}}$ is zero, the edge from $X_k$ to $Y_{k+1}$ vanishes, i.e., $p(X_k=x_k, Y_k=y_k,Y_{k+1} =y_{k+1})= p(X_k=x_k)p(Y_k=y_k|X_k=x_k) p(Y_{k+1}=y_{k+1}|Y_k=y_k)$. (b) Backward transfer entropy $T_{{X^\dagger}_k^{(1)} \to {Y^\dagger}_{k+1}^{(2)}}$ corresponds to the edge from $Y_m$ to $X_{m+1}$ on the Bayesian network. If BTE $T_{{X^\dagger}_k^{(1)} \to {Y^\dagger}_{k+1}^{(2)}}$ is zero, the edge from $Y_m$ to $X_{m+1}$ vanishes, i.e., $p(X_{m+1}=x_{m+1}, Y_m=y_m, Y_{k+1}=y_{k+1})= p(Y_m=y_m) p(Y_{m+1}=y_{m+1}|Y_m=y_m) p(X_{m+1}=x_{m+1}|Y_{m+1}=y_{m+1})$.}
\label{Bayesian}
\end{figure}

{\bf Backward transfer entropy.} Here, we introduce BTE as a novel usage of TE for the backward paths. We first consider the backward path of the system ${\mathcal X}$ (${\mathcal Y}$); ${x^{\dagger}}_{k}^{(l)} := \{x_{N-k+1}, \dots, x_{N-k+l} \}$ (${y^{\dagger}}_{k}^{(l)}=\{y_{N-k+1}, \dots, y_{N-k+l} \}$), which is the time-reversed trajectories of the system $\mathcal X$ ($\mathcal Y$) from time $N-k+l$ to $N-k+1$.
We now introduce the concept of BTE defined as TE for the backward paths
\begin{align} 
T_{{X^{\dagger}}_{k}^{(l)} \to {Y^{\dagger}}_{k'+1}^{(l'+1)}}  &= I({X^{\dagger}}_{k}^{(l)} ;{Y^{\dagger}}_{k'+1}^{(l'+1)}) -I({X^{\dagger}}_{k}^{(l)}; {Y^{\dagger}}_{k'}^{(l')}) = I(X^{(l)}_{m+l} ;Y_{m'} | Y^{(l')}_{m'+l'}),
\label{backward}
\end{align}
with $m = N-k$, $m' =N-k'$ and $k \leq k'$. In this sense, BTE may represent ``the time-reversed directed information flow from the future to the past.'' However BTE is well defined as the conditional mutual information, it is nontrivial if such a concept makes any sense information-theoretically or physically where stochastic dynamics of composite system itself do not necessarily have the time-reversal symmetry. 
 
To clarify the proper meaning of BTE, we compare BTE $T_{{X^\dagger}_k^{(1)} \to {Y^\dagger}_{k+1}^{(2)}}$ with TE $T_{X_k^{(1)} \to Y_{k+1}^{(2)}}$ [see Fig. 1].  
Transfer entropy quantifies the dependence of $X_k$ in the transition from time $Y_k$ to $Y_{k+1}$ [see Fig. 1(a)]. In the same way,  BTE quantifies the dependence of $Y_m$ in the correlation between $X_{m+1}$ and $Y_{m+1}$ [see Fig. 1(b)]. Thus, BTE implies how $X_{m+1}$ depends on $Y_{m+1}$ without the dependence of the past state $Y_{m}$.
In other words,  BTE $T_{{X^\dagger}_k^{(1)} \to {Y^\dagger}_{k+1}^{(2)}}$ is nonnegative and vanishes if and only if a Markov chain $Y_m \to Y_{m+1} \to X_{m+1}$ exists, which implies that dynamics of $X$ are given by a hidden Markov model. In general, BTE $T_{{X^{\dagger}}_{k}^{(l)} \to {Y^{\dagger}}_{k'}^{(l')}}$ is nonnegative and vanishes if and only if a Markov chain 
\begin{align}
&p(X_{m+l}^{(l)} = x_{m+l}^{(l)}, Y_{m'} = y_{m'}, Y^{(l')}_{m' +l'} = y^{(l')}_{m' +l'})  \nonumber \\
=&p( Y_{m'} = y_{m'})p(Y^{(l')}_{m' +l'}  = y^{(l')}_{m' +l'}|Y_{m'} =y_{m'}) p(X_{m+l}^{(l)} = x_{m+l}^{(l)}|Y^{(l')}_{m' +l'}  =y^{(l')}_{m' +l'}),
\label{hiddenM}
\end{align}
exists. Therefore, BTE from $\mathcal X$ to $\mathcal Y$ quantifies how far it is from composite dynamics of $\mathcal X$ and $\mathcal Y$ to a hidden Markov model in $\mathcal X$.

{\bf Thermodynamics of information.} We next discuss a thermodynamic meaning of BTE. To clarify the interpretation of BTE in nonequilibrium stochastic thermodynamics, we consider the following non-Markovian interacting dynamics 
\begin{align}
&p(X_{k+1} = x_{k+1},  Y_{k+1}=y_{k+1} | X_k^{(k)}= x_k^{(k)}, Y_k^{(k)}=  y_{k}^{(k)} ) = P^{\mathcal X}_{k+1} P^{\mathcal Y}_{k+1}, \nonumber\\
&
 P^{\mathcal X}_{k+1} =\begin{cases}
 p(X_{k+1}=x_{k+1} |X_{k}=x_k, Y_{k-n}=y_{k-n} )  & ( k\geq n+1),\\
 p(X_{k+1}=x_{k+1} |X_{k}=x_k , Y_{1}=y_1)  &( k \leq n), 
\end{cases}\nonumber\\
&
P^{\mathcal Y}_{k+1} =\begin{cases}
 p(Y_{k+1}=y_{k+1} |Y_{k}=y_k, X_{k-n}=x_{k-n} )  & ( k\geq n+1),\\
 p(Y_{k+1}=y_{k+1} |Y_{k}=y_k , X_{1}=x_1)  &( k \leq n), 
\end{cases}
\label{nonMarkov}
\end{align}
where a nonnegative integer $n$ represents the time delay between $\mathcal X$ and $\mathcal Y$.
The stochastic entropy change in heat bath $\mathcal B$ attached to the system $\mathcal X$ from time $1$ to $N$ in the presence of $\mathcal Y$~\cite{ItoPRL} is defined as
\begin{align}
\Delta s_{\mathcal B} &:=  \sum_{k=1}^{N-1} \ln \frac{P^{\mathcal X}_{k+1} }{Q_{k+1}^{\mathcal X} } 
\label{DB}
\end{align}
where 
\begin{align}
Q_{k+1}^{X} =\begin{cases}
p_{\rm B}(X_k = x_{k} |X_{k+1} =x_{k+1}, Y_{k-n} = y_{k-n}) & ( k\geq n+1),\\
p_{\rm B} (X_k= x_{k} |X_{k+1}= x_{k+1}, Y_k = y_1) &( k \leq n), 
\end{cases}\nonumber\\
\end{align}
 is the transition probability of backward dynamics, which satisfies the normalization of the probability $\sum_{x_k} Q_{k+1}^{X}  =1$. For example, if the system $\mathcal X$ and $\mathcal Y$ does not include any odd variable that changes its sign with the time-reversal transformation, the backward probability is given by $p_{\rm B} (X_k = x_{k} |X_{k+1} =x_{k+1}, Y_{k-n} = y_{k-n}) =  p(X_{k+1}=x_{k} |X_{k}=x_{k+1}, Y_{k-n}=y_{k-n} )$ with $k\geq n+1$ ($p_{\rm B} (X_k= x_{k} |X_{k+1}= x_{k+1}, Y_k = y_1) =  p(X_{k+1}=x_{k} |X_{k}=x_{k+1} , Y_{1}=y_1) $ with $k\leq n$). 
 This definition of the entropy change in the heat bath Eq. (\ref{DB}) is well known as the local detailed balance or the detailed fluctuation theorem~\cite{Seifert}.
We define the entropy change in $\mathcal X$ and heat bath as
\begin{align}
\Delta S_{\mathcal XB} := \sum_{x_N^{(N)}, y_N^{(N)} }p(X_N^{(N)}=x_N^{(N)}, Y_N^{(N)}=y_N^{(N)}) [\Delta s_{\mathcal B} + \Delta s_{\mathcal X}],
\end{align}
where $ \Delta s_{\mathcal X} := -\ln p(X_N=x_N) + \ln p(X_1 =x_1)$ is the stochastic Shannon entropy change in $\mathcal X$. 

For the non-Markovian interacting dynamics Eq.~(\ref{nonMarkov}), we have the following inequality (see Method);
\begin{align}
 - \Delta S_{\mathcal XB} &\leq  -\sum_{k=1}^{n} \left[T_{{X^\dagger}_{1}^{(1)} \to {Y^\dagger}_{k+1}^{(k+1)}} - T_{X_{1}^{(1)} \to Y_{k+1}^{(k+1)}} \right]  - \sum_{k=n+1}^{N-1} [T_{{X^\dagger}_{k-n}^{(1)} \to {Y^\dagger}_{k+1}^{(k+1)}} - T_{X_{k-n}^{(1)} \to Y_{k+1}^{(k+1)}} ] -I(X_N; Y_N)+ I(X_1; Y_1)
 \label{generalized second law}\\
 &\leq \sum_{k=1}^{n}  T_{X_{1}^{(1)} \to Y_{k+1}^{(k+1)}}  + \sum_{k=n+1}^{N-1}  T_{X_{k-n}^{(1)} \to Y_{k+1}^{(k+1)}}+ I(X_1; Y_1) 
 \label{generalized second law2}.
\end{align}
We add that the term $\sum_{k=1}^{n} \left[T_{{X^\dagger}_{1}^{(1)} \to {Y^\dagger}_{k+1}^{(k+1)}} - T_{X_{1}^{(1)} \to Y_{k+1}^{(k+1)}} \right] $ vanishes for the Markovian interacting dynamics ($n=0$).

These results [Eqs.~(\ref{generalized second law}) and (\ref{generalized second law2})] can be interpreted as a generalized second law of thermodynamics for the subsystem $\mathcal X$ in the presence of information flow from $\mathcal  X$ to $\mathcal  Y$. If there is no interaction between $\mathcal  X$ and $\mathcal Y$, informational terms vanish, i.e., $T_{{X^\dagger}_{1}^{(1)} \to {Y^\dagger}_{k+1}^{(k+1)}} =0$, $T_{X_{1}^{(1)} \to Y_{k+1}^{(k+1)}}=0$, $T_{X_{k-n}^{(1)} \to Y_{k+1}^{(k+1)}}  =0$, $ T_{X_{k-n}^{(1)} \to Y_{k+1}^{(k+1)}}=0$, $I(X_N; Y_N)=0$ and $I(X_1; Y_1)=0$. Thus these results reproduce the conventional second law of thermodynamics $ \Delta S_{\mathcal XB} \geq 0$, which indicates the nonnegativity of the entropy change in $\mathcal X$ and bath~\cite{Seifert}. If there is some interaction between $\mathcal X$ and $\mathcal Y$, $ \Delta S_{\mathcal XB}$ can be negative, and its lower bound is given by the sum of TE from $X$ to $Y$ and mutual information between $\mathcal X$ and $\mathcal Y$ at initial time;
\begin{equation}
I^n( X^{(N)}_{N} \to Y^{(N)}_{N} ):= 
\left\{
\begin{array}{ll}
     I(X_1; Y_1) + \sum_{k=1}^{n}  T_{X_{1}^{(1)} \to Y_{k+1}^{(k+1)}}  + \sum_{k=n+1}^{N-1}  T_{X_{k-n}^{(1)} \to Y_{k+1}^{(k+1)}} & (n \geq 1 )\\
      I(X_1; Y_1) +  \sum_{k=1}^{N-1}  T_{X_{k}^{(1)} \to Y_{k+1}^{(k+1)}} & (n=0)
 \end{array}
\right. ,
\end{equation}
which is a nonnegative quantity $I^n( X^{(N)}_{N} \to Y^{(N)}_{N} ) \geq 0$. In information theory, this quantity $I^0( X^{(N)}_{N} \to Y^{(N)}_{N} )$ is known as DI from $\mathcal X$ to $\mathcal Y$~\cite{Massey}. Intuitively speaking, $- \Delta S_{\mathcal XB}$ quantifies a kind of thermodynamic benefit because its negativity is related to the work extraction in $\mathcal X$ in the presence of $\mathcal Y$~\cite{review}. Thus, a weaker bound (\ref{generalized second law2}) implies that the sum of TE quantifies a possible thermodynamic benefit of $\mathcal X$ in the presence of $\mathcal Y$.

We next consider the sum of TE for the time-reversed trajectories;
\begin{equation}
I^n({X^\dagger}^{(N)}_{N} \to {Y^\dagger}^{(N)}_{N}) = 
\left\{
\begin{array}{ll}
      I(X_N; Y_N)  + \sum_{k=1}^{n} T_{{X^\dagger}_{1}^{(1)} \to {Y^\dagger}_{k+1}^{(k+1)}} + \sum_{k=n+1}^{N-1} T_{{X^\dagger}_{k-n}^{(1)} \to {Y^\dagger}_{k+1}^{(k+1)}} & (n \geq 1 )\\
     I(X_N; Y_N) +  \sum_{k=1}^{N-1} T_{{X^\dagger}_{k}^{(1)} \to {Y^\dagger}_{k+1}^{(k+1)}} & (n=0)
 \end{array}
\right. ,
\end{equation}
which is given by the sum of BTE and the mutual information between $\mathcal X$ and $\mathcal Y$ at final time. A tighter bound (\ref{generalized second law}) can be rewritten as the difference between the sum of TE and BTE;
\begin{align}
- \Delta S_{\mathcal XB} \leq  I^n( X^{(N)}_{N} \to Y^{(N)}_{N} ) - I^n({X^\dagger}^{(N)}_{N} \to {Y^\dagger}^{(N)}_{N}) \leq I^n( X^{(N)}_{N} \to Y^{(N)}_{N} ).
\label{Thermodynamics}
\end{align}
This result implies that a possible benefit $ I^n( X^{(N)}_{N} \to Y^{(N)}_{N} )$ should be reduced by up to the sum of BTE $I^n({X^\dagger}^{(N)}_{N} \to {Y^\dagger}^{(N)}_{N}) $. Thus, the sum of BTE means a loss of thermodynamic benefit. We add that a tighter bound $I^n( X^{(N)}_{N} \to Y^{(N)}_{N} ) - I^n({X^\dagger}^{(N)}_{N} \to {Y^\dagger}^{(N)}_{N})$ is not necessarily nonnegative while a weaker bound $I^n( X^{(N)}_{N} \to Y^{(N)}_{N} )$ is nonnegative.

We here consider the case of Markovian interacting dynamics ($n=0$).  For Markovian interacting dynamics, we have the following additivity for a tighter bound [see Supplementary information (SI)]
\begin{align}
I^0( X^{(N)}_{N} \to Y^{(N)}_{N} ) - I^0({X^\dagger}^{(N)}_{N} \to {Y^\dagger}^{(N)}_{N})  =  \sum_{k=1}^{N-1} \left[ I^0 (X^{(2)}_{k+1} \to Y^{(2)}_{k+1} ) -  I^0( {X^\dagger}^{(2)}_{N-k+1} \to {Y^\dagger}^{(2)}_{N-k+1}) \right],
\label{Markovjump}
\end{align}
where  the sum of TE and BTE for a single time step $I^0 (X^{(2)}_{k+1} \to Y^{(2)}_{k+1} )$ and $I^0( {X^\dagger}^{(2)}_{N-k+1} \to {Y^\dagger}^{(2)}_{N-k+1})$ are defined as $I^0 (X^{(2)}_{k+1} \to Y^{(2)}_{k+1} ) := I(X_k; Y_k) + T_{X^{(1)}_{k} \to Y^{(2)}_{k+1}}$ and $I^0( {X^\dagger}^{(2)}_{N-k+1} \to {Y^\dagger}^{(2)}_{N-k+1}) := I(X_{k+1}; Y_{k+1}) + T_{{X^\dagger}^{(1)}_{N-k} \to {Y^\dagger}^{(2)}_{N-k+1}}$, respectively. 
 This additivity
implies that a tighter bound for multi time steps is equivalent to the sum of a tighter bound for a single time step $I^0 (X^{(2)}_{k+1} \to Y^{(2)}_{k+1} ) -  I^0( {X^\dagger}^{(2)}_{N-k+1} \to {Y^\dagger}^{(2)}_{N-k+1})= I(X_k; \{Y_k, Y_{k+1}\}) - I(X_{k+1}; \{Y_k, Y_{k+1}\}) $. We stress that a tighter bound for a single time step has been derived in Ref.~\cite{ItoNatcom}. We next consider the continuous limit $x_k = x(t= k\Delta t)$, $y_k = y(t= k\Delta t)$, and $N=O(\Delta t^{-1})$, where $t$ denotes continuous time, $\Delta t \ll 1$ is an infinitesimal time interval and the symbol $O$ is the Landau notation. Here we clarify the relationship between a tighter bound (\ref{Markovjump}) and DIF~\cite{HorowitzEsposito} (or the learning rate~\cite{HartichSeifert}) defined as $I_{{\rm flow}}^k := I(X_{k+1} : Y_k) -I(X_k : Y_k)$. For the bipartite Markov jump process~\cite{HartichSeifert} or two dimensional Langevin dynamics without any correlation between thermal noises in $X$ and $Y$~\cite{ItoPRL}, we have the following relationship [see also SI]
\begin{align}
I^0 (X^{(2)}_{k} \to Y^{(2)}_{k+1} ) -  I^0( {X^\dagger}^{(2)}_{N-k+1} \to {Y^\dagger}^{(2)}_{N-k+1}) &= - I_{{\rm flow}}^k + O(\Delta t^2)
\label{BTEflow}.
\end{align}
Thus a bound by TE and BTE is equivalent to a bound by DIF for such systems in the continuous limit, i.e., $- \Delta S_{\mathcal XB} \leq  I^0( X^{(N)}_{N} \to Y^{(N)}_{N} ) - I^0({X^\dagger}^{(N)}_{N} \to {Y^\dagger}^{(N)}_{N}) = \sum_{k=1}^{N-1} I_{{\rm flow}}^k +O(\Delta t)$.

{\bf Gambling with side information. }
In classical information theory, the formalism of the gambling with side information has been well known as another perspective of information theory based on the data compression over a noisy communication channel~\cite{Kelly, Cover-Thomas}. In the gambling with side information, the mutual information between the result in the gambling and the side information gives a bound of the gambler's benefit. 

This formalism of gambling is similar to the above-mentioned result in thermodynamics of information. In thermodynamics, thermodynamic benefit (e.g., the work extraction) can be obtained by using information. On the other hand, the gambler obtain the benefit by using side information. We here clarify the analogy between gambling and thermodynamics in the presence of information flow. To clarify the analogy between thermodynamics and gambling, BTE plays a crucial role as well as TE.

We introduce the basic concept of the gambling with side information given by the horse race~\cite{Kelly, Cover-Thomas}. Let $y_k$ be the horse that won the $k$-th horse race. Let $f_k \geq 0$ and $o_k \geq 0$ be the bet fraction and the odds on the $k$-th race, respectively. Let $m_k$ be the gambler's wealth before the $k$-th race. Let $s_k$ be the side information at time $k$. We consider the set of side information $x_{k-1} = \{s_1, \dots, s_{k-1} \}$, which the gambler can access before the $k$-th race. The bet fraction $f_k$ is given by the function $f_k (y_k| y_{k-1}^{(k-1)}, x_{k-1})$ with $k\geq 2$, and $f_1(y_1|x_1)$. The conditional dependence $\{ y_{k-1}^{(k-1)}, x_{k-1} \}$ ($\{ x_{1} \}$) of $ f_k (y_k| y_{k-1}^{(k-1)}, x_{k-1})$ ($f(y_1|x_1)$) implies that the gambler can decide the bet fraction $f_k$ ($f_1$) by considering the past information $\{ y_{k-1}^{(k-1)}, x_{k-1} \}$ ($\{ x_{1} \}$). We assume normalizations of the bet fractions $\sum_{y_k} f_k(y_k|y_{k-1}^{(k-1)}, x_{k-1}) =1$ and $\sum_{y_1} f_1(y_1|x_1) =1$, which mean that the gambler bets all one's money in every race. We also assume that $\sum_{y_k} 1/o_k (y_k)=1$. This condition satisfies if the odds in every race are fair, i.e., $1/o_k (y_k)$ is given by a probability of $Y_k$.

The stochastic gambler's wealth growth rate at $k$-th race is given by
\begin{align}
g_k := \ln \frac{m_{k+1}}{m_k} =\ln[ f_k (y_k| y_{k-1}^{(k-1)}, x_{k-1}) o_k (y_k)],
\end{align}
with $k\geq 2$ [$g_1 :=\ln (m_2/m_1) = f_1(y_1|x_1) o_{1} (y_1)$], which implies that the gambler's wealth stochastically changes due to the bet fraction and odds. The information theory of the gambling with side information indicates that the ensemble average of total wealth growth $G := \sum_{x_N^{(N)}, y_N^{(N)}} p (X_N^{(N)} =x_N^{(N)}, Y_N^{(N)} =y_N^{(N)} )[\sum_{k=1}^{N} g_k]$ is bounded by the sum of TE (or DI) from $X$ to $Y$~\cite{Permuter, Hidaka} (see Method);
\begin{align}
G \leq& \sum_{k=1}^{N}\langle  \ln o_k \rangle  -S (Y_N^{(N)}) +I^0( X^{(N)}_{N} \to Y^{(N)}_{N} )
 \label{Gamble}\\
\leq& I^0( X^{(N)}_{N} \to Y^{(N)}_{N} ),
\label{Gamble2}
\end{align}
where $\langle  \cdots \rangle =\sum_{x_N^{(N)}, y_N^{(N)}} p (X_N^{(N)} =x_N^{(N)}, Y_N^{(N)} =y_N^{(N)} ) \cdots$ indicates the ensemble average, and $S(Y_N^{(N)} ) := - \langle \ln p (Y_N^{(N)} =y_N^{(N)} ) \rangle$ is the Shannon entropy of $Y_N^{(N)}$.
This result (\ref{Gamble2}) implies that the sum of TE can be interpreted as a possible benefit of the gambler. 

\begin {figure}
\centering
\includegraphics[width=7cm]{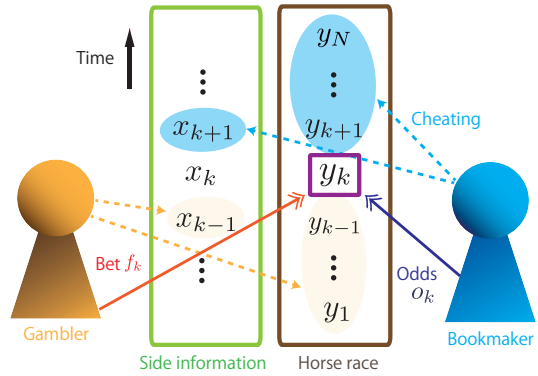}
\caption{Schematic of the special case of the horse race. The gambler can only access the past side information $x_{k-1}$ and the past races $y^{(k-1)}_{k-1} = \{y_1, \dots, y_{k-1} \}$, and decides the bet fraction $f_k$ on the $k$-th race. The bookmaker makes some cheating which can access the future side information $x_{k+1}$ and the future races $y^{(N-k)}_{N} = \{y_{k+1}, \dots, y_{N} \}$, and decides the odds on the $k$-th race.}
\label{Ga}
\end{figure}

We discuss the analogy between thermodynamics of information and the gambling with side information. A weaker bound in the gambling with side information (\ref{Gamble2}) is similar to a weaker bound in thermodynamics of information (\ref{Thermodynamics}), where the negative entropy change $-\Delta S_{\mathcal XB}$ corresponds to the total wealth growth $G$. On the other hand, a tighter bound in the gambling with side information (\ref{Gamble}) is rather different from a tighter bound by the sum of BTE in thermodynamics of information (\ref{Thermodynamics}). We show that a tighter bound in the gambling is also given by the sum of BTE if we consider the special case that the bookmaker who decides the odds $o_k$ cheats in the horse race; The odds $o_k$ can be decided by the unaccessible side information $x_{k+1}$ and information of the future races $y_{N}^{(N-k)}$  [see also Fig.~\ref{Ga}]. In this special case, the fair odds of the $k$-th race $o_k$ can be the conditional probability of the future information $1/o_k (y_k) = p(Y_k= y_k|Y_N^{(N-k)} = y_{N}^{(N-k)}, X_{k+1} =x_{k+1})$ with $k \leq N-1$, and $1/o_N (y_N) = p(Y_N= y_N| X_N= x_{N})$. The inequality (\ref{Gamble}) can be rewritten as
\begin{align}
G \leq  I^0( X^{(N)}_{N} \to Y^{(N)}_{N} ) - I^0({X^\dagger}^{(N)}_{N} \to {Y^\dagger}^{(N)}_{N}) \leq I^0( X^{(N)}_{N} \to Y^{(N)}_{N} ).
\label{G-Ineq} 
\end{align}
which implies that the sum of BTE $I^0( X^{(N)}_{N} \to Y^{(N)}_{N} ) - I^0({X^\dagger}^{(N)}_{N} \to {Y^\dagger}^{(N)}_{N}$ represents a loss of the gambler's benefit because of the cheating by the bookmaker who can access the future information with anti-causality. We stress that Eq.~(\ref{G-Ineq}) has a same form of the thermodynamic inequality~(\ref{Thermodynamics}) for Markovian interacting dynamics ($n=0$). This fact implies that thermodynamics of information can be interpreted as the special case of the gambling with side information; The gambler uses the past information and the bookmaker uses the future information. If we regard thermodynamic dynamics as the gambling, anti-causal effect should be considered.

{\bf Causality.} We here show that BTE itself is related to anti-causality without considering the gambling.
From the view point of the statistical hypothesis test, TE is equivalent to GC for Gaussian variables~\cite{BarnettSeth}. Therefore, it is naturally expected that BTE can be interpreted as a kind of the causality test.

Suppose that we consider two linear regression models
\begin{align}
y_{k'+1}^{(1)} &= \alpha+  (y_{k'}^{(l')} \oplus  x_{k}^{(l)}) \cdot A + \epsilon,\\
y_{k'+1}^{(1)} &= \alpha' + (y_{k'}^{(l')} ) \cdot A' + \epsilon',
\end{align}
where $\alpha$ ($\alpha'$) is a constant term, $A$ ($A'$) is the vector of regression coefficients, $\oplus$ denotes concatenation of vectors, and $\epsilon$ ($ \epsilon'$) is an error term. The Granger causality of $\mathcal X$ to $\mathcal Y$ quantifies how the past time series of $\mathcal X$ in the first model reduces the prediction error of $y_{k'+1}^{(1)}$ compared to the error in the second model.  Performing ordinary mean squares to find the regression coefficients $A$ ($A'$) and $\alpha$ ($\alpha'$) that minimize the variance of $\epsilon$ ($\epsilon'$), the standard measure of GC is given by 
\begin{align}
{\mathcal F}_{X_{k}^{(l)} \to Y_{k'+1}^{(l' +1)}} := \ln \frac{{\rm var}(\epsilon')}{ {\rm var}(\epsilon)},
\end{align}
where ${\rm var} (\epsilon)$ denotes the variance of $\epsilon$. Here we assume that the joint probability $p(X_{k}^{(l)} = x_k^{(l)}, Y_{k'+1}^{(l'+1)} = y_{k'+1}^{(l'+1)})$ is Gaussian. Under Gaussian assumption, TE and GC are equivalent up to a factor of $2$,
\begin{align}
2 T_{X_{k}^{(l)} \to Y_{k'+1}^{(l' +1)}}  = {\mathcal F}_{X_{k}^{(l)} \to Y_{k'+1}^{(l' +1)}}.
\end{align}
In the same way, we discuss BTE from the view point of GC. Here we assume that the joint probability $p(X_{m+l}^{(l)}= x_{m+l}^{(l)}, Y_{m'+l'}^{(m'+l')} =y^{(l'+1)}_{m' +l'})$ is Gaussian. Suppose that two linear regression models
\begin{align}
{y^\dagger}_{k'+1}^{(1)}&= \alpha^{\dagger}+ ({y^\dagger}_{k'}^{(l')} \oplus  {x^\dagger}_{k}^{(l)}) \cdot A^{\dagger} + \epsilon^{\dagger} ,\\
{y^\dagger}_{k'+1}^{(1)} &= \alpha'^{\dagger} + ({y^\dagger}_{k'}^{(l')} ) \cdot A'^{\dagger} + \epsilon'^{\dagger},
\end{align}
where $\alpha^{\dagger}$ ($\alpha'^{\dagger}$) is a constant term, $A^{\dagger}$ ($A'^{\dagger}$) is the vector of regression coefficients and $\epsilon^{\dagger}$ ($ \epsilon'^{\dagger}$) is an error term. These linear regression models give a prediction of the past state of $\mathcal Y$ using the future time series of $\mathcal X$ and $\mathcal Y$. Intuitively speaking, we consider GC of $\mathcal X$ to $\mathcal Y$ for the rewind playback video of composite dynamics $\mathcal X$ and $\mathcal Y$. We call this causality test the Granger {\it anti-}causality of $\mathcal X$ to $\mathcal Y$. Performing ordinary mean squares to find $A^{\dagger}$ ($A'^{\dagger}$) and $\alpha^{\dagger}$ ($\alpha'^{\dagger}$)  that minimize ${\rm var} (\epsilon^{\dagger})$ ($ {\rm var}  (\epsilon'^{\dagger })$) , we define a measure of the Granger  {\it anti-}causality of $\mathcal X$ to $\mathcal Y$ as ${\mathcal F}_{{X^\dagger}_{k}^{(l)} \to {Y^\dagger}_{k'+1}^{(l' +1)}} := \ln [{\rm var}(\epsilon'^{\dagger}) / {\rm var}(\epsilon^{\dagger}) ]$. The backward transfer entropy is equivalent to the Granger  {\it anti-}causality up to factor $2$, 
\begin{align}
2 T_{{X^\dagger}_{k}^{(l)} \to {Y^\dagger}_{k'+1}^{(l' +1)}}  = {\mathcal F}_{{X^\dagger}_{k}^{(l)} \to {Y^\dagger}_{k'+1}^{(l' +1)}}.
\label{anticausal}
\end{align}
This fact implies that BTE can be interpreted as a kind of {\it anti-}causality test. We stress that composite dynamics of $\mathcal X$ and $\mathcal Y$ are not necessarily driven with {\it anti-}causality even if a measure of the Granger {\it anti-}causality $ {\mathcal F}_{{X^\dagger}_{k}^{(l)} \to {Y^\dagger}_{k'+1}^{(l' +1)}}$ has nonzero value. As GC just finds only the predictive causality~\cite{Grangerbooks,Granger}, the Granger {\it anti-}causality also finds only the predictive causality for the backward time series.

{\bf Discussion} 

We proposed that directed measure of information called BTE, which is possibly useful to detect a hidden Markov model (\ref{hiddenM}) and predictive anti-causality (\ref{anticausal}). In the both setting of thermodynamics and the gambling, the measurement of BTE has a profitable meaning; the detection of a loss of a possible benefit in the inequalities (\ref{Thermodynamics}) and (\ref{G-Ineq}). 

The concept of BTE can provide a clear perspective in the studies of the biochemical sensor and thermodynamics of information, because the difference between TE and DIF has attracted attention recently in these fields~\cite{HartichSeifertSensor, HorowitzSandberg}. In Ref.~\cite{HartichSeifertSensor}, Hartich {\it et al.} have proposed the novel informational measure for the biochemical sensor called {\it sensory capacity}. The sensory capacity is defined as the ratio between TE and DIF $C := -I_{{\rm flow}}^k/ T_{X_k^{(1)} \to Y_{k+1}^{(2)}}$. Because DIF can be rewritten by TE and BTE [Eq. (\ref{BTEflow})] for Markovian interacting dynamics, we have the following expression for the sensory capacity in a stationary state,
\begin{align}
C = 1 -\frac{T_{{X^\dagger}_{N-k}^{(1)} \to {Y^\dagger}_{N-k+1}^{(2)} } }{T_{X_{k}^{(1)} \to Y_{k+1}^{(2)} }}, 
\label{sensor}
\end{align}
where we used $I(X_{k+1}; Y_{k+1}) = I(X_{k}; Y_{k}) $ in a stationary state. This fact indicates that the ratio between TE and BTE could be useful to quantify the performance of the biochemical sensor. By using this expression (\ref{sensor}), we show that the maximum value of the sensory capacity $C=1$ can be achieved if a Markov chain of a hidden Markov model $Y_k \to Y_{k+1} \to X_{k+1}$ exists. In Ref.~\cite{HorowitzSandberg}, Horowitz and Sandberg have shown a comparison between two thermodynamic bound by TE and DIF for two dimensional Langevin dynamics. For the Kalman-Bucy filter which is the optimal controller, they have found the fact that DIF is equivalent to TE in a stationary state. This idea can be clarified by the concept of BTE. Because the Kalman-Bucy filter can be interpreted as a hidden Markov model, BTE should be zero, and DIF is equivalent to TE in a stationary state. 

Our results can be interpreted as a generalization of previous works in thermodynamics of information~\cite{Still, Still2, Diana}. In Refs.~\cite{Still, Still2}, S. Still {\it et al.} discuss the prediction in thermodynamics for Markovian interacting dynamics. In our results, we show the connection between thermodynamics of information and the predictive causality from the view point of GC. Thus, our results give a new insight into these works of the prediction in thermodynamics. In Ref.~\cite{Diana}, G. Diana and M. Esposito have introduced the time-reversed mutual information for Markovian interacting dynamics. In our results, we introduce BTE, which is TE in the time-reversed way. Thus, our result provides a similar description of thermodynamics by introducing BTE, even for non-Markovian interacting dynamics.

We point out the time symmetry in the generalized second law (\ref{generalized second law}). For Markovian interacting dynamics, the equality in Eq. (\ref{generalized second law}) holds if dynamics of $\mathcal X$ has a local reversibility (see SI). Here we consider a time reversed transformation $\mathcal{T}: k \to N-k+1$, and assume a local reversibility such that the backward probability $p_{\rm B}(A=a|B=b)$ equals to the original probability $p(A=a|B=b)$ for any random variables $A$ and $B $. In a time reversed transformation, we have $\mathcal{T}: \Delta S_{\mathcal XB} \to -  \Delta S_{\mathcal XB}$,  $\mathcal{T}: I^n (X^{(N)}_{N} \to Y^{(N)}_{N} ) \to I^n ({X^\dagger}^{(N)}_{N} \to {Y^\dagger}^{(N)}_{N} )$ and $\mathcal{T}: I^n ({X^\dagger}^{(N)}_{N} \to {Y^\dagger}^{(N)}_{N} ) \to I^n ( X^{(N)}_{N} \to Y^{(N)}_{N} ) $. The generalized second law Eq.~(\ref{generalized second law}) changes the sign in a time reversed transformation, 
$\mathcal{T}:  \left[ 0 \leq \Delta S_{\mathcal XB} +I^n (X^{(N)}_{N} \to Y^{(N)}_{N} ) - I^n ({X^\dagger}^{(N)}_{N} \to {Y^\dagger}^{(N)}_{N} )  \right]  \to \left[ 0 \leq - \left( \Delta S_{\mathcal XB}  + I^n (X^{(N)}_{N} \to Y^{(N)}_{N} ) -I^n ({X^\dagger}^{(N)}_{N} \to {Y^\dagger}^{(N)}_{N} ) \right) \right]$.
Thus, the generalized second law (\ref{generalized second law}) has the same time symmetry in the conventional second law, i.e., $\mathcal{T}: [0 \leq \Delta S_{\rm tot}] \to [0 \leq -\Delta S_{\rm tot} ]$ even for non-Markovian interacting dynamics, where $\Delta S_{\rm tot}$ is the entropy change in total systems. In other words, the generalized second law (\ref{generalized second law}) provides the arrow of time as the conventional second law. 
This fact may indicate that BTE is useful as well as TE in physical situations where the time symmetry plays a crucial role in physical laws.

We also point out that this paper clarifies the analogy between thermodynamics of information and the gambling. The analogy between the gambling and thermodynamics has been proposed in Ref.~\cite{Dror}, however, the analogy between Eqs.~(\ref{Thermodynamics}) and ~(\ref{G-Ineq}) are different from one in Ref.~\cite{Dror}. In Ref.~\cite{Dror}, D. A. Vinkler {\it et al.} discuss the particular case of the work extraction in Szilard engine, and consider the work extraction in Szilard engine as the gambling. On the other hand, our result provides the analogy between the general law of thermodynamics of information and the gambling.
To clarify this analogy, we may apply the theory of gambling, for example the portfolio theory~\cite{Cover,Permuter}, to thermodynamic situations in general.
We also stress that the gambling with side information directly connects with the data compression in information theory~\cite{Cover-Thomas}. Therefore, the generalized second law of thermodynamics may directly connect with the data compression in information theory. To consider such applications, BTE would play a tricky role in the theory of the gambling where the odds should be decided with anti-causality.

Finally, we discuss the usage of BTE in time series analysis. In principle, we prepare the backward time series data from the original time series data, and do a calculation of BTE as TE. To calculate BTE, we can estimate how far it is from dynamics of two time series to a hidden Markov model, or detect the predictive causality for the backward time series. In physical situations, we also can detect thermodynamic performance by comparing BTE with TE. If the sum of BTE from the target system to the other systems is larger than the sum of TE from the target system to the other systems, the target system could seem to violate the second law of thermodynamics because of the inequality (\ref{Thermodynamics}), where the other systems play a similar role of Maxwell's demon. Therefore, BTE could be useful to detect phenomena of Maxwell's demon in several settings such as Brownian particles~\cite{Toyabe, Berut}, electric devices~\cite{Koski, Aki}, and biochemical networks~\cite{ItoNatcom, Sartori, Infopross, Bo, tenWolde, McGrath}.

{\bf Method}

{\bf The outline of the derivation of inequality (\ref{generalized second law}).}  
We here show the outline of the derivation of the generalized second law (\ref{generalized second law}) [see also SI for details]. In SI, we show that the quantity $ \Delta S_{\mathcal XB}+ I^n (X^{(N)}_N \to Y^{(N)}_N ) - I^n ({X^\dagger}^{(N)}_N \to {Y^\dagger}^{(N)}_N )$, can be rewritten as the Kullbuck-Leiber divergence $D_{\rm KL} (\rho|| \tilde{\rho}) :=  \sum_{x_N^{(N)}, y_{N}^{(N)}} \rho(x_N^{(N)}, y_{N}^{(N)}) \ln [\rho(x_N^{(N)}, y_{N}^{(N)})/\tilde{\rho}(x_N^{(N)}, y_{N}^{(N)}) ]$~\cite{Cover-Thomas}, where $\rho (x_N^{(N)}, y_{N}^{(N)}):= p(X_N^{(N)} =x_N^{(N)}, Y_N^{(N)} =y_{N}^{(N)})$ and $\tilde{\rho}  (x_N^{(N)}, y_{N}^{(N)}) := p(X_N= x_N, Y_N= y_N) \prod_{k'=n+1}^{N-1}  \! p_{\rm B} (X_{k'} = x_{k'} |X_{k'+1} = x_{k'+1}, Y_{k'-n} =y_{k'-n} )\! \prod_{m'=N-n}^{N-1} \! p( Y_{m'}=y_{m'}|Y_{m'+1}=y_{m'+1}, X_{N}=x_{N} ) \! \prod_{k=1}^{n}\! p_{\rm B}(X_k= x_{k} | X_{k+1}=x_{k+1}, Y_1= y_1) \! \prod_{m=1}^{N-n-1} \! p( Y_m= y_{m}|Y_{m+1} =y_{m+1},X_{m+n+1} = x_{m+n+1} )$, are nonnegative functions that satisfy the normalizations $\sum_{x_N^{(N)}, y_{N}^{(N)}} \rho(x_N^{(N)}, y_{N}^{(N)})= 1$ and $\sum_{x_N^{(N)}, y_{N}^{(N)}} \tilde{\rho}(x_N^{(N)}, y_{N}^{(N)}) =1$. Due to the nonnegativity of the Kullbuck-Leiber divergence, we obtain the inequality (\ref{generalized second law}), i.e., $ \Delta S_{\mathcal XB}+ I^n (X^{(N)}_N \to Y^{(N)}_N ) - I^n ({X^\dagger}^{(N)}_N \to {Y^\dagger}^{(N)}_N ) \geq 0$. We add that the integrated fluctuation theorem corresponding to the inequality (\ref{generalized second law}) is also valid, i.e., $\sum \rho(x_N^{(N)}, y_{N}^{(N)}) \exp ( -\ln [\rho(x_N^{(N)}, y_{N}^{(N)})/\tilde{\rho}(x_N^{(N)}, y_{N}^{(N)}) ]) =1$. 

{\bf The outline of the derivation of inequality (\ref{Gamble}).}  
We here show the outline of the derivation of the gambling inequality (\ref{Gamble}) [see also SI for details]. The quantity $-G + \sum_{k=1}^N \langle  \ln o_k \rangle -S (Y_N^{(N)})  +I^0 (X^{(N)}_N \to Y^{(N)}_N )  $ can be rewritten as the Kullbuck-Leiber divergence $D_{\rm KL} (\rho|| \pi) $,  $\rho (x_N^{(N)}, y_{N}^{(N)}):= p(X_N^{(N)} =x_N^{(N)}, Y_N^{(N)} =y_{N}^{(N)})$ and $\pi  (x_N^{(N)}, y_{N}^{(N)}) :=  f_1(y_1|x_1) p(X_1=x_1) \prod_{k=1}^{N-1} p(X_{k+1}=x_{k+1}| Y_k^{(k)}=y_k^{(k)}, X_k= x_k) f_{k+1}(y_{k+1}|y_{k}^{(k)}, x_{k})$, are nonnegative functions that satisfy the normalizations $\sum_{x_N^{(N)}, y_{N}^{(N)}} \rho(x_N^{(N)}, y_{N}^{(N)})= 1$ and $\sum_{x_N^{(N)}, y_{N}^{(N)}} \pi(x_N^{(N)}, y_{N}^{(N)}) =1$.
 Due to the nonnegativity of the Kullbuck-Leiber divergence, we have the inequality  (\ref{Gamble}), i.e., $-G + \sum_{k=1}^{N} \langle  \ln o_k \rangle - S (Y_N^{(N)}) + I^0 (X^{(N)}_N \to Y^{(N)}_N ) \geq 0 $.
		
\subsection{Acknowledgements}	

We are grateful to Takumi Matsumoto, Takahiro Sagawa, Naoto Shiraishi and Shumpei Yamamoto for the fruitful discussion on a range of issues of thermodynamics.  This work was supported by Grant-in-Aid for JSPS Fellows No.~27-7404 and by JSPS KAKENHI Grant No.~16752084.

\subsection*{Supplementary Information}
{\bf Supplementary note 1 $|$ Detailed derivation of the inequality (12) in the main text.}
To simplify the calculation, we use the notation $\rho(a) :=p(A=a)$ and $\rho(a|b) := p(A=a|B=b)$ for any random variables $A$ and $B$. We also use the notation $\rho_B (a|b) :=p_B (A=a|B=b)$ for any random variables $A$ and $B$. We define ensemble average as $\langle \cdots \rangle = \sum_{x^{(N)}_N, y^{(N)}_N} \rho(x^{(N)}_N, y^{(N)}_N) \cdots$. We here consider the following non-Markovian interacting dynamics,
 \begin{align}
\rho(x_N^{(N)}, y_N^{(N)} ) :=  \rho(x_1, y_1) \prod_{k'=1}^{n} \rho (x_{k'+1} | x_{k'}, y_1) \rho(y_{k'+1}|y_{k'}, x_1) \prod_{k=n+1}^{N-1} \rho (x_{k+1} | x_{k}, y_{k -n }) \rho(y_{k+1} | y_{k}, x_{k-n }),
\end{align}
with $n \geq 1$, and 
 \begin{align}
\rho(x_N^{(N)}, y_N^{(N)} ) :=  \rho(x_1, y_1)  \prod_{k=1}^{N-1} \rho (x_{k+1} | x_{k}, y_{k}) \rho(y_{k+1} | y_{k}, x_{k}),
\end{align}
with $n=0$.

To derive the inequality (12) in the main text, we calculate the difference between $I^n( X^{(N)}_{N} \to Y^{(N)}_{N} ) - I^n({X^\dagger}^{(N)}_{N} \to {Y^\dagger}^{(N)}_{N})$ and $-\Delta S_{\mathcal XB}$ in the case of $n \geq 1$ as follows;
 \begin{align}
&\Delta S_{\mathcal XB} + I^n( X^{(N)}_{N} \to Y^{(N)}_{N} ) - I^n({X^\dagger}^{(N)}_{N} \to {Y^\dagger}^{(N)}_{N})  \nonumber\\
=&\Delta S_{\mathcal XB}   -\sum_{k=1}^{n} \left[T_{{X^\dagger}_{1}^{(1)} \to {Y^\dagger}_{k+1}^{(k+1)} }- T_{X_{1}^{(1)} \to Y_{k+1}^{(k+1)} } \right]  - \sum_{k=n+1}^{N-1} \left[T_{{X^\dagger}_{k-n}^{(1)} \to {Y^\dagger}_{k+1}^{(k+1)}} - T_{X_{k-n}^{(1)} \to Y_{k+1}^{(k+1)}} \right] -I(X_N; Y_N)+ I(X_1; Y_1)  \nonumber \\ 
=& \left< \ln \left[ \frac{\rho(x_1|y_1) }{\rho (x_N|y_N) } \prod_{k'=1}^{n} \frac{ \rho(x_{k'+1} | x_{k'}, y_1) }{\rho_{\rm B}(x_{k'} | x_{k'+1}, y_1)  } \prod_{k=n+1}^{N-1} \frac{ \rho(x_{k+1} | x_{k}, y_{k-n} ) }{\rho_{\rm B} (x_{k} | x_{k+1}, y_{k-n} )} \right] \right> + \left< \ln  \left[ \prod_{k'=1}^{n} \frac{\rho( y_{k'+1}|y_{k'} ,x_1) }{\rho( y_{k'+1}|y_{k'}^{(k')}) }  \prod_{k=n+1}^{N-1} \frac{\rho( y_{k+1}|y_{k} ,x_{k-n}) }{\rho( y_{k+1}|y_{k}^{(k)}) } \right] \right> \nonumber\\
& +\left< \ln \prod_{k'=1}^{n} \frac{ \rho( y_{N-k'}|y^{(k')}_{N} ) }{ \rho( y_{N-k'}|x_{N} ,y_{N-k'+1})}   \prod_{k=n+1}^{N-1} \frac{ \rho( y_{N-k}|y^{(k)}_{N} ) }{ \rho( y_{N-k}|x_{N-k+n+1} ,y_{N-k+1})} \right> \nonumber \\
=& \left< \ln \frac{ \! \rho(x_1, y_1) \!\prod_{k'=1}^{n}\! \rho (x_{k'+1} | x_{k'}, y_1) \rho(y_{k'+1}|y_{k'}, x_1) \! \prod_{k=n+1}^{N-1} \! \rho (x_{k+1} | x_{k}, y_{k -n }) \rho(y_{k+1} | y_{k}, x_{k-n })}{\! \rho(x_N, y_N) \! \prod_{k'=n+1}^{N-1}  \! \rho_{\rm B} (x_{k'} | x_{k'+1}, y_{k'-n} )\! \prod_{m'=N-n}^{N-1} \! \rho( y_{m'}|y_{m'+1}, x_{N} ) \! \prod_{k=1}^{n}\! \rho_{\rm B}(x_{k} | x_{k+1}, y_1) \! \prod_{m=1}^{N-n-1} \! \rho( y_{m}|y_{m+1}, x_{m+n+1} ) }  \right> \nonumber \\
=& \sum_{x_N^{(N)}, y_N^{(N)}} \rho(x_N^{(N)}, y_N^{(N)} ) \ln \frac{\rho(x_N^{(N)}, y_N^{(N)} ) }{\tilde{\rho}(x_N^{(N)}, y_N^{(N)} ) },
\label{app-1}
\end{align}
where we used $\rho( y_{k'+1}|y_{k'}^{(k')} ,x_1)  =\rho( y_{k'+1}|y_{k'} ,x_1)$ with $k' \leq n$, $\rho( y_{k+1}|y_{k}^{(k)} ,x_{k-n})  =\rho( y_{k+1}|y_{k} ,x_{k-n})$ with $k \geq n+1$, $\rho( y_{N-k'}|x_{N} ,y_N^{(k')})=\rho( y_{N-k'}|x_{N} ,y_{N-k'+1})$ with $k' \leq n$, $ \rho( y_{N-k}|x_{N-k+n+1} ,y^{(k)}_N)= \rho( y_{N-k}|x_{N-k+n+1} ,y_{N-k+1})$ with $k \geq n+1$, and $\tilde{\rho}(x_N^{(N)}, y_N^{(N)}) $ is defined as
 \begin{align}
\tilde{\rho}(x_N^{(N)}, y_N^{(N)} ):= \! \rho(x_N, y_N) \! \prod_{k'=n+1}^{N-1}  \! \rho_{\rm B} (x_{k'} | x_{k'+1}, y_{k'-n} )\! \prod_{m'=N-n}^{N-1} \! \rho( y_{m'}|y_{m'+1}, x_{N} ) \! \prod_{k=1}^{n}\! \rho_{\rm B}(x_{k} | x_{k+1}, y_1) \! \prod_{m=1}^{N-n-1} \! \rho( y_{m}|y_{m+1}, x_{m+n+1} ).
\end{align}

In the case of $n=0$, we have
 \begin{align}
&\Delta S_{\mathcal XB} + I^0( X^{(N)}_{N} \to Y^{(N)}_{N} ) - I^0({X^\dagger}^{(N)}_{N} \to {Y^\dagger}^{(N)}_{N})  \nonumber\\
=&\Delta S_{\mathcal XB}  - \sum_{k=1}^{N-1} \left[T_{{X^\dagger}_{k}^{(1)} \to {Y^\dagger}_{k+1}^{(k+1)}} - T_{X_{k}^{(1)} \to Y_{k+1}^{(k+1)}} \right] -I(X_N; Y_N)+ I(X_1; Y_1)  \nonumber \\ 
=& \left< \ln \left[ \frac{\rho(x_1|y_1) }{\rho (x_N|y_N) } \prod_{k=1}^{N-1} \frac{ \rho(x_{k+1} | x_{k}, y_{k} ) }{\rho_{\rm B} (x_{k} | x_{k+1}, y_{k} )} \right] \right> + \left< \ln  \left[ \prod_{k=1}^{N-1} \frac{\rho( y_{k+1}|y_{k} ,x_{k}) }{\rho( y_{k+1}|y_{k}^{(k)}) } \right] \right> +\left< \ln \prod_{k=1}^{N-1} \frac{ \rho( y_{N-k}|y^{(k)}_{N} ) }{ \rho( y_{N-k}|x_{N-k+1} ,y_{N-k+1})} \right> \nonumber \\
=& \left< \ln \frac{ \! \rho(x_1, y_1)  \prod_{k=1}^{N-1} \! \rho (x_{k+1} | x_{k}, y_{k}) \rho(y_{k+1} | y_{k}, x_{k})}{\! \rho(x_N, y_N) \! \prod_{k=1}^{N-1}  \rho_{\rm B} (x_{k} | x_{k+1}, y_{k} )  \prod_{k=1}^{N-1} \! \rho( y_{k}|y_{k+1}, x_{k+1} ) }  \right> \nonumber \\
=& \sum_{x_N^{(N)}, y_N^{(N)}} \rho(x_N^{(N)}, y_N^{(N)} ) \ln \frac{\rho(x_N^{(N)}, y_N^{(N)} ) }{\tilde{\rho}(x_N^{(N)}, y_N^{(N)} ) },
\label{app-2}
\end{align}
where we used $\rho( y_{k+1}|y_{k}^{(k)} ,x_{k})  =\rho( y_{k+1}|y_{k} ,x_{k})$, $ \rho( y_{N}|x_{N-k+1} ,y^{(k)}_N)= \rho( y_{N-k}|x_{N-k+1} ,y_{N-k+1})$, and $\tilde{\rho}(x_N^{(N)}, y_N^{(N)} )$ is defined as
 \begin{align}
\tilde{\rho}(x_N^{(N)}, y_N^{(N)} ):= \rho(x_N, y_N) \! \prod_{k=1}^{N-1}  \rho_{\rm B} (x_{k} | x_{k+1}, y_{k} )  \prod_{m=1}^{N-1} \! \rho( y_{m}|y_{m+1}, x_{m+1} ) .
\end{align}

The function $\tilde{\rho}(x_N^{(N)}, y_N^{(N)} )$ is nonnegative, and satisfies the normalization of the probability;
 \begin{align}
&\sum_{x_N^{(N)}, y_N^{(N)}} \tilde{\rho}(x_N^{(N)}, y_N^{(N)} )\nonumber\\
 &= \sum_{x_N^{(N)}, y_N^{(N)}}  \! \rho(x_N, y_N) \! \prod_{k'=n+1}^{N-1}  \! \rho_{\rm B} (x_{k'} | x_{k'+1}, y_{k'-n} )\! \prod_{m'=N-n}^{N-1} \! \rho( y_{m'}|y_{m'+1}, x_{N} ) \! \prod_{k=1}^{n}\! \rho_{\rm B}(x_{k} | x_{k+1}, y_1) \! \prod_{m=1}^{N-n-1} \! \rho( y_{m}|y_{m+1}, x_{m+n+1} )\nonumber\\
 &= \sum_{x_N^{(N-n)}, y_N^{(N)}}  \! \rho(x_N, y_N) \! \prod_{k'=n+1}^{N-1}  \! \rho_{\rm B} (x_{k'} | x_{k'+1}, y_{k'-n} )\! \prod_{m'=N-n}^{N-1} \! \rho( y_{m'}|y_{m'+1}, x_{N} ) \! \prod_{m=1}^{N-n-1} \! \rho( y_{m}|y_{m+1}, x_{m+n+1} )
\nonumber\\
 &= \sum_{x_N^{(N-n-1)}, y_N^{(N-1)}}   \! \rho(x_N, y_N) \! \prod_{k'=n+2}^{N-1}  \! \rho_{\rm B} (x_{k'} | x_{k'+1}, y_{k'-n} )\! \prod_{m'=N-n}^{N-1} \! \rho( y_{m'}|y_{m'+1}, x_{N} ) \! \prod_{m=2}^{N-n-1} \! \rho( y_{m}|y_{m+1}, x_{m+n+1} ) \nonumber\\
  &= \cdots\nonumber\\
  &= \sum_{x_N^{(1)}, y_N^{(n+1)}}  \rho(x_N, y_N) \!  \prod_{m'=N-n}^{N-1} \! \rho( y_{m'}|y_{m'+1}, x_{N} )  \nonumber\\
    &= \sum_{x_N^{(1)}, y_N^{(1)}}  \rho(x_N, y_N)  \nonumber\\
    &= 1,
    \end{align}
with $n \geq 1$, and
 \begin{align}
&\sum_{x_N^{(N)}, y_N^{(N)}} \tilde{\rho}(x_N^{(N)}, y_N^{(N)} )\nonumber\\
 &= \sum_{x_N^{(N)}, y_N^{(N)}}  \! \rho(x_N, y_N) \! \prod_{k=1}^{N-1}  \! \rho_{\rm B} (x_{k} | x_{k+1}, y_{k} ) \prod_{m=1}^{N-1} \! \rho( y_{m}|y_{m+1}, x_{m+1} )\nonumber\\
 &= \sum_{x_N^{(N-1)}, y_N^{(N-1)}}   \! \rho(x_N, y_N) \prod_{k=2}^{N-1}  \! \rho_{\rm B} (x_{k} | x_{k+1}, y_{k} ) \prod_{m=2}^{N-1} \! \rho( y_{m}|y_{m+1}, x_{m+1} ) \nonumber\\
  &= \cdots\nonumber\\
    &= \sum_{x_N^{(1)}, y_N^{(1)}}  \rho(x_N, y_N)  \nonumber\\
    &= 1,
    \end{align}
with $n=0$.

    Thus Eqs. (\ref{app-1}) and (\ref{app-2}) are given by the Kullback-Libler divergence between $\rho(x_N^{(N)}, y_N^{(N)} )$ and $\tilde{\rho}(x_N^{(N)}, y_N^{(N)} )$. Because of the nonnegativity of the Kullback-Libler divergence~\cite{SupCover-Thomas}, we have Eq.~(12) in the main text
\begin{align}
- \Delta S_{\mathcal XB}  \leq  -\sum_{k=1}^{n} \left[T_{{X^\dagger}_{1}^{(1)} \to {Y^\dagger}_{k+1}^{(k+1)} }- T_{X_{1}^{(1)} \to Y_{k+1}^{(k+1)} } \right]  - \sum_{k=n+1}^{N-1} \left[T_{{X^\dagger}_{k-n}^{(1)} \to {Y^\dagger}_{k+1}^{(k+1)}} - T_{X_{k-n}^{(1)} \to Y_{k+1}^{(k+1)}} \right] -I(X_N; Y_N)+ I(X_1; Y_1),
\end{align}
for $n \geq 1$, and
\begin{align}
- \Delta S_{\mathcal XB}  \leq  - \sum_{k=1}^{N-1} \left[T_{{X^\dagger}_{k}^{(1)} \to {Y^\dagger}_{k+1}^{(k+1)}} - T_{X_{k}^{(1)} \to Y_{k+1}^{(k+1)}} \right] -I(X_N; Y_N)+ I(X_1; Y_1) ,
\label{Sup-Markov}
\end{align}
for $n=0$.
The equality holds if and only if $\rho(x_N^{(N)}, y_N^{(N)} )= \tilde{\rho}(x_N^{(N)}, y_N^{(N)} )$. In the case of $n=0$, this condition is given by
 \begin{align}
\rho(x_1, y_1)  \prod_{k=1}^{N-1} \rho (x_{k+1} | x_{k}, y_{k}) \rho(y_{k+1} | y_{k}, x_{k}) &=\rho(x_N, y_N) \! \prod_{k=1}^{N-1}  \rho_{\rm B} (x_{k} | x_{k+1}, y_{k} )  \prod_{m=1}^{N-1} \! \rho( y_{m}|y_{m+1}, x_{m+1} ) \\
\rho(x_N, y_N)  \prod_{k=1}^{N-1} \rho (x_{k}, y_{k} | y_{k+1}, x_{k+1}) &=\rho(x_N, y_N) \! \prod_{k=1}^{N-1}  \rho_{\rm B} (x_{k} | x_{k+1}, y_{k} )  \prod_{m=1}^{N-1} \! \rho( y_{m}|y_{m+1}, x_{m+1} )  \\
\prod_{k=1}^{N-1} \rho (x_{k}| y_{k}, y_{k+1}, x_{k+1}) &=\prod_{k=1}^{N-1}  \rho_{\rm B} (x_{k} | x_{k+1}, y_{k} ),
\end{align}
which implies the backward probability $\rho_{\rm B} (x_{k} | x_{k+1}, y_{k} )$ is equivalent to the original probability $\rho (x_{k}| y_{k}, y_{k+1}, x_{k+1})$. In a continuous limit, this fact implies the equality in the generalized second law (\ref{Sup-Markov}) holds when the dynamics of $\mathcal X$ has a local reversibility, i.e., $\rho =\rho_B $.

{\bf Supplementary note 2 $|$ Detailed calculation of Eqs. (17) and (18) in the main text.}

We consider the following Markovian interacting dynamics
 \begin{align}
\rho(x_N^{(N)}, y_N^{(N)}) = \rho(x_1, y_1) \prod_{k=1}^{N-1} \rho(x_{k+1} | x_{k}, y_{k}) \rho(y_{k+1} | x_{k}, y_{k}).
\label{appB}
 \end{align}
 From Eq. (\ref{appB}), we have $\rho(y_k|y_N^{(N-k)}, x_{k+1} ) =\rho(y_k|y_{k+1}, x_{k+1} )$ and $\rho(y_{k+1}|y_{k}^{(k)}, x_{k} )=\rho(y_{k+1}|y_{k}, x_{k} )$. We also have an identity $\rho(y_k|y_{k+1}) \rho(y_{k+1}) =\rho(y_{k+1}|y_{k}) \rho(y_{k}) $. 
Thus we can calculate the additivity Eq.~(17) in the main text as follows;
\begin{align}
&I^0( X^{(N)}_{N} \to Y^{(N)}_{N} ) - I^0({X^\dagger}^{(N)}_{N} \to {Y^\dagger}^{(N)}_{N}) \nonumber\\
&= \sum_{k=1}^{N-1} [T_{X_{k}^{(1)} \to Y_{k+1}^{(k+1)}} - T_{{X^\dagger}_{k}^{(1)} \to {Y^\dagger}_{k+1}^{(k+1)}}] +I(X_1; Y_1) -I(X_N; Y_N) \nonumber\\
&= I(X_1; Y_1) +\sum_{k=1}^{N-1} I(X_k; Y_{k+1} |Y_{k}^{(k)})  -I(X_N; Y_N) - \sum_{k=1}^{N-1} I(X_{k+1}; Y_{k} | Y_{N}^{(N-k)})\nonumber \\
&= \left< \ln \left[ \frac{\rho(y_N)\rho(y_1|x_1)}{\rho(y_1)\rho(y_N|x_N)} \prod^{N-1}_{k=1} \frac{\rho(y_k|y_N^{(N-k)} )}{p(y_k|y_N^{(N-k)}, x_{k+1} )}  \frac{p(y_{k+1}|y_{k}^{(k)}, x_{k} )}{p(y_{k+1}|y_{k}^{(k)} )} \right] \right>\nonumber \\
&= \left< \ln  \left[ \frac{ \rho(y_1|x_1)}{\rho(y_N|x_N)} \prod^{N-1}_{k=1} \frac{\rho(y_k|y_{k+1} ) \rho(y_{k+1})}{p(y_k|y_{k+1}, x_{k+1} )}  \frac{\rho(y_{k+1}|y_k, x_{k} )}{\rho(y_{k+1}|y_{k} ) \rho(y_{k})} \right]  \right>\nonumber \\
&= I(X_1; Y_1) +\sum_{k=1}^{N-1} I(X_k; Y_{k+1} |Y_{k})  -I(X_N; Y_N) - \sum_{k=1}^{N-1} I(X_{k+1}; Y_{k} | Y_{k+1} ) \nonumber \\
&= \sum_{k=1}^{N-1} [I(X_k; Y_{k+1} | Y_{k}) +I(X_{k}; Y_{k}) -I(X_{k+1}; Y_{k} | Y_{k+1}) - I(X_{k+1}; Y_{k+1}) ]\nonumber \\
&= \sum_{k=1}^{N-1} \left[ I^0 (X^{(2)}_{k+1} \to Y^{(2)}_{k+1} ) -  I^0( {X^\dagger}^{(2)}_{N-k+1} \to {Y^\dagger}^{(2)}_{N-k+1}) \right].
\end{align}

The difference between a tighter bound and DIF is calculated as follows;
\begin{align}
&I^0 (X^{(2)}_{k+1} \to Y^{(2)}_{k+1} ) -  I^0( {X^\dagger}^{(2)}_{N-k+1} \to {Y^\dagger}^{(2)}_{N-k+1})  + I_{\rm flow}^k  \nonumber\\
&=  I(X_{k}; \{Y_k, Y_{k+1} \})- I(X_{k+1}; \{Y_k, Y_{k+1} \}) - I(X_{k}; Y_k) +I(X_{k+1}; Y_k)\nonumber \\
&=  I(X_{k}; Y_{k+1}|Y_k )- I(X_{k+1};  Y_{k+1} |Y_k)  \nonumber \\
&=  \left< \ln \left[ \frac{\rho(y_{k+1} | y_k, x_k )}{\rho(y_{k+1} | y_k, x_{k+1} )} \right] \right>
\end{align}
For the bipartite Markov jump process~\cite{SupHartichSeifert} or two dimensional Langevin dynamics without any correlation between thermal noises in $\mathcal X$ and $\mathcal Y$~\cite{SupItoPRL}, the ratio between two transition rates in $\mathcal Y$,  i.e., $\langle  \ln [\rho(y_{k+1}| y_{k}, x_{k}) /\rho(y_{k+1}| y_{k}, x_{k+1} )] \rangle$ is up to order $O(\Delta t^2)$.

{\bf Supplementary note 3 $|$ Comparison between a tighter bound in Eq. (16) and the result in [Ito, S., \& Sagawa, T., Phys. Rev. Lett. {\bf 111}, 180603 (2013)].}

We compare a tighter bound in Eqs. (16) with our previous result in Ref.~\cite{SupItoPRL}. For the non-Markovian interacting dynamics
 \begin{align}
\rho(x_N^{(N)}, y_N^{(N)} ) :=  \rho(x_1, y_1) \prod_{k'=1}^{n} \rho (x_{k'+1} | x_{k'}, y_1) \rho(y_{k'+1}|y_{k'}, x_1) \prod_{k=n+1}^{N-1} \rho (x_{k+1} | x_{k}, y_{k -n }) \rho(y_{k+1} | y_{k}, x_{k-n }),
\end{align}
with $n\geq 1$, the previous result in Ref.~\cite{SupItoPRL} gives the following bound of the entropy change in $\mathcal X$ and bath,
\begin{align}
 - \Delta S_{\mathcal XB} &\leq  -\langle \Theta \rangle  \nonumber \\
 &:= I(X_1; Y_1) +\sum_{k=1}^{n}  T_{X_{1}^{(1)} \to Y_{k+1}^{(k+1)} } +\sum_{k=n+1}^{N-1} T_{X_{k-n}^{(1)} \to y_{k+1}^{(k+1)}} - I(X_N; Y_N^{(N)}) \nonumber\\
 &:= I^n(X_{N}^{(N)} \to Y_{N}^{(N)}) - I(X_N; Y_N) - \sum_{k=1}^{N-1} T_{{X^\dagger}_{1}^{(1)} \to {Y^\dagger}_{k+1}^{(k+1)} } \nonumber\\
 \label{PRL2013}
\end{align}
where we used the identity $I(X_N; Y_N^{(N)})= I(X_N; Y_N) + \sum_{k=1}^{N-1} T_{{X^\dagger}_{1}^{(1)} \to {Y^\dagger}_{k+1}^{(k+1)} }$.  Here, $ I(X_1; Y_1)$ corresponds to the initial correlation term $I_{\rm ini}$, $\sum_{k=1}^{n}  T_{X_{1}^{(1)} \to Y_{k+1}^{(k+1)} } +\sum_{k=n+1}^{N-1} T_{X_{k-n}^{(1)} \to y_{k+1}^{(k+1)}}$ corresponds the transfer entropy term $\sum_l I_{\rm tr}^l$, and $I(X_N; Y_N^{(N)})$ corresponds to the final correlation term $I_{\rm fin}$ in Ref.~\cite{SupItoPRL}. The difference between $-\langle \Theta \rangle $ and $ I^n( X^{(N)}_{N} \to Y^{(N)}_{N} ) - I^n({X^\dagger}^{(N)}_{N} \to {Y^\dagger}^{(N)}_{N})$ can be calculated as the difference of BTE,
\begin{align}
  -\langle \Theta \rangle  - [ I^n( X^{(N)}_{N} \to Y^{(N)}_{N} ) - I^n({X^\dagger}^{(N)}_{N} \to {Y^\dagger}^{(N)}_{N})] =  - \sum_{k=n+1}^{N-1} [T_{{X^\dagger}_{1}^{(1)} \to {Y^\dagger}_{k+1}^{(k+1)} } -T_{{X^\dagger}_{k-n}^{(1)} \to {Y^\dagger}_{k+1}^{(k+1)}} ].
  \end{align}
Due to the conditional Markov chain 
\begin{align}
\rho(x_N ,  x_{N-k +n+1}, y_{N-k}| {y}_N^{(k)} ) = \rho(  x_N |x_{N-k +n+1}, {y}_N^{(k)} )\rho( x_{N-k +n+1}|y_{N-k}, {y}_N^{(k)} ) \rho(  y_{N-k}| {y}_N^{(k)} ),
\end{align}
we have the data processing inequality–\cite{SupCover-Thomas}
\begin{align}
T_{{X^\dagger}_{1}^{(1)} \to {Y^\dagger}_{k+1}^{(k+1)} } = I( X_N ;Y_{N-k} |Y_N^{(k)})  \leq   I( X_{N-k+n+1} ; Y_{N-k} |Y_N^{(k)}) = T_{{X^\dagger}_{k-n}^{(1)} \to {Y^\dagger}_{k+1}^{(k+1)}}.
\end{align}
 Therefore, a tighter bound in inequality Eq. (16) is tighter than a bound in the previous result~\cite{SupItoPRL}, 
\begin{align} 
 - \Delta S_{\mathcal XB} \leq I^n( X^{(N)}_{N} \to Y^{(N)}_{N} ) - I^n({X^\dagger}^{(N)}_{N} \to {Y^\dagger}^{(N)}_{N}) \leq   -\langle \Theta \rangle.
  \end{align}

{\bf Supplementary note 4 $|$ Detailed derivation of the inequality (20) in the main text.}

The set of side information $x_k$ satisfies $x_{k-1} = \{s_1, \dots, s_k \} \subset x_{k}$. Thus we have $\rho(y_{k+1}|y_{k}^{(k)} , x_k^{(k)}) = \rho(y_{k+1}|y_{k}^{(k)} , x_k)$ and $\rho(x_{k+1}| y_k^{(k)}, x^{(k)}_k)= \rho(x_{k+1}| y_k^{(k)}, x_k)$. The joint probability is given by 
 \begin{align}
\rho(x_N^{(N)}, y_N^{(N)}) = \rho(x_1, y_1) \prod_{k=1}^{N-1} \rho(x_{k+1}| y_k^{(k)}, x_k) \rho(y_{k+1}|y_{k}^{(k)} , x_k).
\label{appC}
 \end{align}
Thus, we can calculate as follows;
\begin{align}
&-G+ \sum_{k=1}^{N}\langle  \ln o_k \rangle  -S (Y_N^{(N)}) +I^0( X^{(N)}_{N} \to Y^{(N)}_{N} )\nonumber \\
&=- \langle  \ln f_1 (y_{1}|x_1) \rangle  - \sum_{k=1}^{N-1}\langle  \ln f_{k+1} (y_{k+1}|y_{k}^{(k)}, x_k) \rangle  + \langle \ln \rho (y_N^{(N)}) \rangle + \left< \ln  \left[ \frac{\rho( x_{1}, y_{1}) }{\rho(x_1) \rho(y_1) } \right] \right> + \left< \ln  \left[ \prod_{k=1}^{N-1} \frac{\rho( y_{k+1}|y_{k}^{(k)},x_{k}) }{\rho( y_{k+1}|y_{k}^{(k)}) } \right] \right>  \nonumber \\
&= \left< \ln \left[\frac{ \rho(x_1, y_1)}{f_1(y_1|x_1) \rho(x_1)} \prod_{k=1}^{N-1}  \frac{ \rho(y_{k+1}|y_{k}^{(k)} , x_k)}{ f_{k+1}(y_{k+1}|y_{k}^{(k)}, x_{k})} \right]+ \ln \frac{\rho(y^{(N)}_N)}{\rho(y^{(N)}_N)}  \right> \nonumber\\
&= \left< \ln \left[\frac{ \rho(x_1, y_1)}{f_1(y_1|x_1) \rho(x_1)} \prod_{k=1}^{N-1}  \frac{ \rho(x_{k+1}| y_k^{(k)}, x_k)  \rho(y_{k+1}|y_{k}^{(k)} , x_k)}{  \rho(x_{k+1}| y_k^{(k)}, x_k) f_{k+1}(y_{k+1}|y_{k}^{(k)}, x_{k}) } \right]  \right> \nonumber\\
&= \sum_{x^{(N)}_N, y^{(N)}_N} \rho(x^{(N)}_N, y^{(N)}_N) \ln \frac{\rho(x^{(N)}_N, y^{(N)}_N)}{\pi(x^{(N)}_N, y^{(N)}_N)},
\label{app-C}
\end{align}
where $\pi(x^{(N)}_N, y^{(N)}_N)$ is defined as
\begin{align}
\pi(x^{(N)}_N, y^{(N)}_N) := f_1(y_1|x_1) \rho(x_1) \prod_{k=1}^{N-1} \rho(x_{k+1}| y_k^{(k)}, x_k) f_{k+1}(y_{k+1}|y_{k}^{(k)}, x_{k}) . 
\end{align}
The function $\pi(x_N^{(N)}, y_N^{(N)} )$ satisfies the normalization of the probability,
\begin{align}
\sum_{x_N^{(N)}, y_N^{(N)}} \pi(x_N^{(N)}, y_N^{(N)} ) &= \sum_{x_N^{(N)}, y_N^{(N)}}  f_1(y_1|x_1) \rho(x_1) \prod_{k=1}^{N-1}f_{k+1}(y_{k+1}|y_{k}^{(k)}, x_{k}) \rho(x_{k+1}| y_k^{(k)}, x_k) \nonumber \\
&= \sum_{x_{N-1}^{(N-1)}, y_{N-1}^{(N-1)}}  f_1(y_1|x_1) \rho(x_1) \prod_{k=1}^{N-2} f_{k+1}(y_{k+1}|y_{k}^{(k)}, x_{k}) \rho(x_{k+1}| y_k^{(k)}, x_k) \nonumber \\
&= \cdots \nonumber \\
&= \sum_{x_{1}, y_{1}}  f_1(y_1|x_1) \rho(x_1) \nonumber\\
&=\sum_{x_{1}} \rho(x_1) \nonumber\\
&=1,
\end{align}
where we used $\sum_{y_1}  f_1(y_1|x_1)=1$ and $\sum_{y_{k+1}} f_{k+1}(y_{k+1}|y_{k}^{(k)}, x_{k}) =1$.

   Thus Eq. (\ref{app-C}) is the Kullback-Libler divergence between $p(x_N^{(N)}, y_N^{(N)} )$ and $\pi(x_N^{(N)}, y_N^{(N)} )$. Because of the nonnegativity of the Kullback-Libler divergence~\cite{SupCover-Thomas}, we have Eq.~(20) in the main text
\begin{align}
-G+ \sum_{k=1}^{N}\langle  \ln o_k \rangle  -S (Y_N^{(N)}) +I^0( X^{(N)}_{N} \to Y^{(N)}_{N} ) \geq 0.
\end{align}
The equality holds if and only if $p(x_N^{(N)}, y_N^{(N)} )= \pi(x_N^{(N)}, y_N^{(N)} )$. This condition is given by
\begin{align}
 \rho(x_1, y_1) \prod_{k=1}^{N-1} \rho(x_{k+1}| y_k^{(k)}, x_k) \rho(y_{k+1}|y_{k}^{(k)} , x_k) &= f_1(y_1|x_1) \rho(x_1) \prod_{k=1}^{N-1} \rho(x_{k+1}| y_k^{(k)}, x_k) f_{k+1}(y_{k+1}|y_{k}^{(k)}, x_{k}) \nonumber\\
  \rho(y_1|x_1) \prod_{k=1}^{N-1}  \rho(y_{k+1}|y_{k}^{(k)} , x_k) &=f_1(y_1|x_1) \prod_{k=1}^{N-1} f_{k+1}(y_{k+1}|y_{k}^{(k)}, x_{k}),
\end{align}
which implies the bet fraction $f_k$ is equivalent to the original probability $\rho$, i.e., the proportional betting is optimal.

\end{document}